\documentclass[epj]{svjour}
\usepackage{epsfig}

\begin{document}

\author{M.N.Kiselev$^{1,2}$, H.Feldmann$^1$ and R.Oppermann$^1$}
\institute{$^1$Institut f\"ur Theoretische Physik, Universit\"at W\"urzburg,
  D-97074 W\"urzburg, Germany\\
  $^2$ Russian Research Center "Kurchatov Institute", 123 182 Moscow,
  Russia}
\date{\today}
\title{Semi-fermionic representation of SU($N$) Hamiltonians}

\abstract{
  We represent the generators of the SU($N$) algebra as bilinear
  combinations of Fermi operators with imaginary chemical potential.
  The distribution function, consisting of a minimal set of discrete
  imaginary chemical potentials, is introduced to satisfy 
  the local constraints.  This
  representation leads to the conventional temperature diagram
  technique with standard Feynman codex, except that the Matsubara
  frequencies are determined by neither integer nor half-integer
  numbers.  The real-time Schwinger-Keldysh formalism is formulated in the
  framework of complex equilibrium distribution functions
  for auxiliary semi-fermionic fields.  We discuss the
  continuous large $N$ and SU(2) large spin limits. We illustrate the
  application of this technique for magnetic and spin-liquid states 
  of the Heisenberg model.
} 
\PACS{75.10.-b, 75.10.Jm, 75.40.Gb, 71.10.Fd}

\titlerunning{Semi-fermionic representation of SU(N) Hamiltonians}
\maketitle

Several approaches have been proposed for the description of spin systems
in statistical physics. Methods of functional integration based on
various representation of spin operators such as Fermi, Bose,
Majorana, supersymmetric or Hubbard operators
\cite{holstein40a,dyson56a,maleev58a,abrikosov65a,abrikosov67a,larkin68a,larkin68b,hubbard65a,coleman00a,coleman00b,coleman00c} 
have been applied to many problems involving
quantum spins and pseudospins \cite{read83a,auerbach86a,millis87a,bicker87a,kotliar86a,sachdev89a,sachdev89b,sachdev90a,affleck88a,affleck89a,auerbach88a,chubukov91a}.  The
difficulty with the representation of spin operators is connected with
the fact that spins possess neither Fermi nor Bose statistics. The
commutation relations for spins are determined by the SU(2) algebra,
leading to the absence of a Wick theorem for SU(2) generators.  The
Gaudin \cite{gaudin60a} theorem existing instead makes it impossible to
construct a simple diagram technique directly for spin operators. To
resolve this problem, various representations \cite{dyson56a,maleev58a,abrikosov65a,abrikosov67a,larkin68a,larkin68b,hubbard65a,coleman00a,coleman00b,coleman00c}
have been introduced.  Nevertheless, the representation of spins as a
bilinear combination of Fermi/Bose operators enlarges the
dimensionality of Hilbert space where these operators act.  Thus, the
spurious (unphysical) states should be excluded from the consideration
resulting in a constraint requirement. Basically, different
representations cure the constraint problem in a different way.
Nevertheless, the usual price for simplicity is the replacement of the
local constraint on each point containing the spins by a so-called
global constraint, so that the restriction is fulfilled only in the 
average over all sites.  It is known that such a replacement results
in uncontrollable approximations for quantum spins (especially in low
dimensions). Although the use of a global constraint is questionable
for SU(2) systems, it becomes more reasonable for higher SU($N$)
groups, especially in the "large $N$ limit". The corresponding
approach is known as "$1/N$ expansion", \cite{sachdev89a,sachdev89b,sachdev90a,affleck88a,affleck89a}
successfully describing the strong coupling limits of the Kondo
impurity \cite{read83a}, Anderson lattice \cite{auerbach86a,millis87a,bicker87a} and
Hubbard \cite{kotliar86a} models and also SU($N$) Heisenberg
antiferromagnets on a square lattice \cite{sachdev89a,sachdev89b,sachdev90a,affleck88a,affleck89a,auerbach88a} shedding light on the
mechanism of high T$_c$ superconductivity in cuprate compounds.
Although SU($N=2$) models are of primary physical interest, the
SU($N\ne2$) models can be considered as "approximate models" where an
"exact solution" can be gained in contrast to "exact models" where the
"approximate solution" is hard to obtain \cite{affleck88a,affleck89a}.
The simplification arises due to expansion in the inverse number of
"flavors" $1/N$, making it possible to start with mean-field solution
and systematically find corrections to it.

The goal of this paper is to consider a semi-fermionic representation
for SU($N$) generators for arbitrary (not necessary large) $N$,
applying a different idea of constraint realization. This idea is know
as Popov-Fedotov \cite{popov88a} representation being initially proposed
for $S=1/2$ and $S=1$ spins. Based on an exact representation of spin
operators as fermions with imaginary chemical potential, this
representation resulted in the conventional Feynman temperature
diagram technique, nevertheless providing a rigorous treatment of the
local constraint.
\cite{oppermann91a,gros90a,kiselev99a,kiselev00a,kiselev00b,oppermann99b}
In this paper we give a generalization of this
method to SU($N$) and we also construct the real-time
formalism for the semi-fermionic spin representation.

The SU($N$) algebra is determined by the generators obeying the
following commutational relations:
\begin{equation}
[\hat S^\beta_{\alpha, i} \hat S^\rho_{\sigma j}]=
\delta_{ij}(\delta^\rho_\alpha \hat S^\beta_{\sigma i}-\delta^\beta_\sigma
\hat S^\rho_{\alpha i})
\end{equation}
where $\alpha,\beta=1,...,N$. We adopt the definition of the 
Cartan algebra \cite{cartan} of the SU($N$) group
$\{H_\alpha\}=S_\alpha^\alpha$ similar to the one used in \cite{sachdev89a},
noting that the diagonal generators $S_\alpha^\alpha$ are not traceless.
To ensure a vanishing trace, the
diagonal generators should only appear in combinations
\begin{equation}
\sum_{\alpha=1}^{N} s_\alpha S_\alpha^\alpha \quad \mbox{with} \quad
\sum_{\alpha=1}^{N} s_\alpha = 0
\end{equation}
which effectively reduces the number of independent diagonal generators to
$N-1$ and the total number of SU($N$) generators to $N^2-1$.

\begin{figure}
\begin{center}
  \epsfxsize36mm \epsfbox{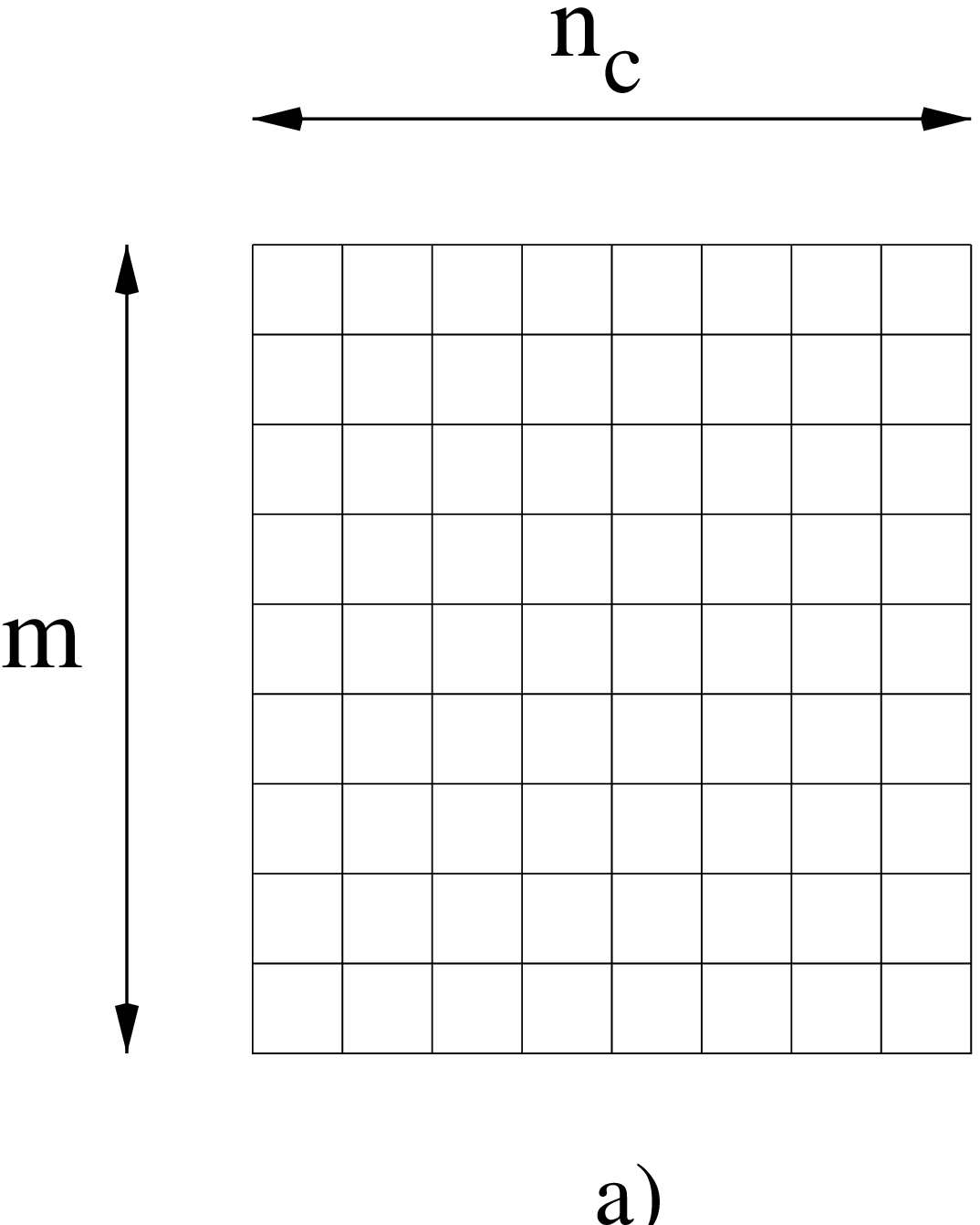}\hspace*{1cm}
  \epsfxsize36mm \epsfbox{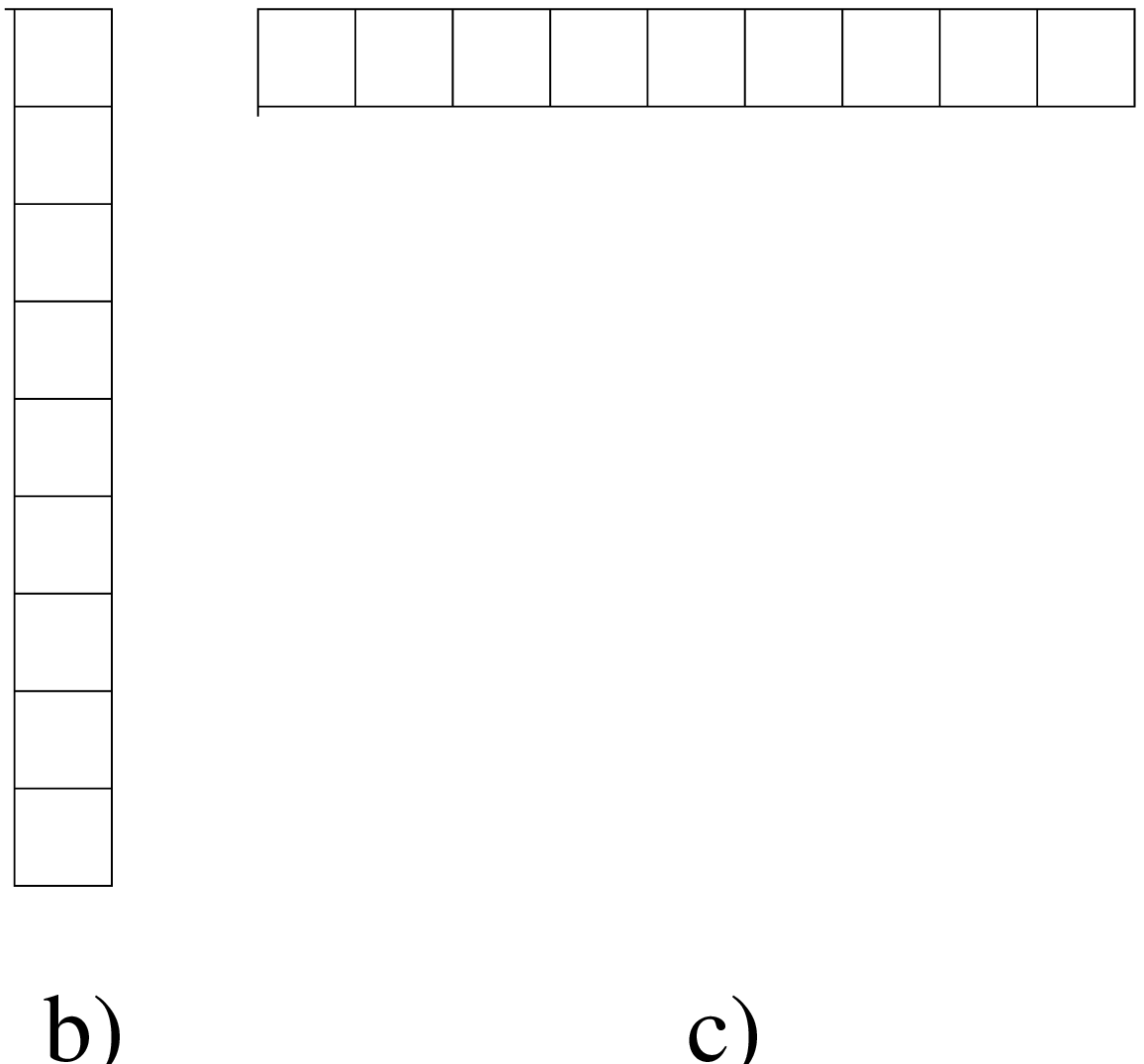}
\end{center}
\caption{
Rectangular  Young tableau to denote a SU($N$) representation (a),
single column tableau corresponding to $n_c=1$ (b) and single
row tableau standing for spin $S=n_c/2$ representation of SU(2) group (c).}
\label{fig:young_tableaux}
\end{figure}

For SU(2) one recognizes the usual spin operators
\begin{equation}
S_1^2=S^+,\quad S_2^1=S^-,\quad S_2^2-S_1^1=2S^z
\end{equation}
with the usual commutation relations \cite{auerbach94a} and the Pauli matrices
as generators of the SU(2) group.  We shall not confine ourself to some
special type of Hamiltonian. Nevertheless, it's worthwhile to mention
that the SU($N$) generalization of the Heisenberg model is given by
the following expression \cite{sachdev89a,affleck88a,affleck89a}
\begin{equation}
H= \frac{J}{N}
\sum_{<ij>}\sum_{\alpha\beta}\hat S^\beta_\alpha(i)\hat S^\alpha_\beta(j)
\end{equation}
On each site, there may exist many particles, whose symmetry
properties define a specific representation of SU($N$).  The most
transparent way to visualize an irreducible SU($N$) representation are
Young tableaux \cite{chen89a,lichtenberg70a}.  Instead of the general
Young tableau, specified by $N-1$ integers, for example the lengths of
the rows, we restrict us for the main part of the paper to rectangular
tableaux, with $1 \leq m \leq N$ rows and $n_c \geq 1$ columns,
illustrated in Fig.\ref{fig:young_tableaux}.

The familiar spin is given by $N=2$, $m=1$, $n_c=2S$, so the Young
tableau contains one row of $2S$ length, with only one box for
$S=1/2$.  Also of special importance are the tableaux with $n_c=1$,
giving the $N-1$ fundamental representations of SU($N$).

The $\hat S^\alpha_\beta$ generator may be written as biquadratic form
in terms of Schwinger boson operators \cite{auerbach88a},\cite{auerbach94a}:
\begin{equation}
\hat S^\alpha_\beta=b^\dagger_{\alpha p} b^{\beta p}
\end{equation}
and a constraint as follows
\begin{equation}
\sum_{\alpha=1}^Nb^\dagger_{\alpha p} b^{\alpha q}=\delta_p^q n_c
\end{equation}
where $p=1,...,m$ is the number of "colors".

The equivalent fermionic representation of the generators of
SU($N$)\cite{sachdev89a} is given by
\begin{equation}
\hat S^\alpha_\beta=\sum_ac^\dagger_{\alpha a}c^{\beta a}
\label{fermi}
\end{equation}
where the "color" index $a,b=1,...,n_c$ and the $n_c(n_c+1)/2$
constraints
\begin{equation}
\sum_{\alpha=1}^Nc^\dagger_{\alpha a} c^{\alpha b}=\delta_a^b m
\label{gconst}
\end{equation}
restrict the Hilbert space to the states with $m * n_c$ particles and
ensure the characteristic symmetry in the color index $a$.  The
antisymmetric behavior with respect to $\alpha$ is a direct
consequence of the fermionic representation.

Let us consider the partition function for the Hamiltonian, expressed
in terms of SU($N$) generators
\begin{equation}
Z_{S}=Tr \exp(-\beta H_S)
\end{equation}
For SU(2) $S=1/2$ and $S=1$ it is possible to map the spin partition
function onto a fermionic partition function where the chemical
potential of fermions is purely imaginary \cite{popov88a}
\begin{equation}
Z_S=A \; Tr\exp\left(-\beta(H_F-\mu N_F)\right)
=A Z_F
\end{equation} 
with $\mu=-i\pi T/2$ and $A=i^{n_{site}}$ for $S=1/2$ and $\mu=-i\pi
T/3$ and $A=(i/\sqrt{3})^{n_{site}}$ for $S=1$, $n_{site}$ denotes 
the number of sites in a lattice. This results in usual
Feynman-like diagram technique built up with the help of auxiliary
Fermi (Grassmann) fields. The corresponding Matsubara frequencies for
Popov-Fedotov (PF) fermions after applying the generalized Grassmann
boundary conditions \cite{popov88a} read as $\omega_n=2\pi T(n+1/4)$ for
$S=1/2$ and $\omega_n=2\pi T(n+1/3)$ for $S=1$. The imaginary chemical
potentials are important for the realization of an exclusion principle
providing the fulfillment of the general identity
\begin{equation}
Z_S= Tr \exp(- \beta H_S) = Tr \exp(- \beta H_F) \delta_{n_F,1}
\label{part1}
\end{equation}
The Popov-Fedotov representation has been generalized for arbitrary
values of spin $S$ for the 
SU(2) group in \cite{oppermann94a} by introducing the
distribution of discrete chemical potentials $\mu(j)$, with $j$ being
the site index, for PF fermions:
\begin{equation}
Z_S=\prod_{j}\int d\mu(j) P(\mu(j))Z_F(\mu(j))
\label{part0}
\end{equation}

For the SU($N$) algebra
we shall try to find the partition function in a similar manner
$$
Z_S=\int\prod_{j} d\mu(j) P(\mu(j)) 
Tr\exp\left(-\beta(H_F-\mu(j)n_F)\right)=
$$
\begin{equation}
=\int\prod_{j} d\mu(j) P(\mu(j))Z_F(\mu(j))
\end{equation}
We use the path integral representation of the partition function
$$
Z_S/Z_S^0=\int\prod_{j} d\mu(j) P(\mu(j))\exp({\cal A})/
$$
\begin{equation}
\int\prod_{j} d\mu(j) P(\mu(j))\exp({\cal A}_0)
\end{equation}
where the action ${\cal A}$ and ${\cal A}_0$ are determined by
$$
{\cal A}={\cal A}_0 - \int_0^\beta d\tau H_F(\tau),
$$
\begin{equation}
{\cal A}_0=\sum_{j}\sum_{k=1}^N\int_0^\beta d\tau\bar a_k(j,\tau)
(\partial_\tau+\mu(j))a_k(j,\tau)
\end{equation}
and the fermionic representation of SU($N$) generators (\ref{fermi})
is applied.  

To begin with we confine ourselves to two particular
cases of SU($N$) with $n_c=1$ (corresponding to an effective "spin
size" $S=1/2$ and in the language of Young tableaux described by one
column) and SU(2) for arbitrary value of $n_c=2S$ (one row Young
tableau).

Let us first consider $n_c=1$. We denote the corresponding
distribution by $P_{N,m}(\mu(j))$, where $m$ is the number of particles
in the SU($N$) orbital, or in other words, $1\leq m < N$ labels the
different fundamental representations of SU($N$).
\begin{equation}
n_j=\sum_{k=1}^N \bar a_{k}(j) a_{k}(j)=m
\label{const1}
\end{equation}
To satisfy this requirement, the minimal set of chemical potentials
and the corresponding form of $P_{N,m}(\mu(j))$ are to be derived.

Let us classify the states in Fock and spin spaces.  We note that the
dimension of the Fock space is $dim H_F=2^N$ and spurious states
should be excluded.  Thus, there are $\nu(N,m)=C_N^{m}=N!/(m!(N-m)!)$
physical states which can be obtained from the vacuum state
$\Phi_0=\underbrace{|0,...,0\rangle}_{N}$ as follows
\begin{equation}
\Phi^{\{\nu\}}_{phys}=(\prod_{l=1}^{m}a^\dagger_l) \Phi_0
\end{equation}
or from
the$|\underbrace{1,...,1}_{m},\underbrace{0,...,0}_{N-m}\rangle$ state
by transferring the occupied states from left to the right side using
the group generators.

To derive the distribution function we use the following identity for
constraint (\ref{const1}) expressed in terms of Grassmann variables
\begin{equation}
\delta_{n_j, m}=\frac{1}{N}
\sin\left(\pi(n_j-m)\right)/
\sin\left(\frac{\pi(n_j-m)}{N}\right)
\label{d1}
\end{equation}
Substituting this identity into (\ref{part1}) and comparing with
(\ref{part0}) on gets
\begin{equation}
P_{N,m}(\mu(j))=\frac{1}{N}\sum_{k=1}^{N}
\exp\left(\frac{i\pi m}{N}(2k-1)\right)\delta(\mu(j)-\mu_k)
\label{eq:P_v1}
\end{equation}
where
\begin{equation}
\mu_k = - \frac{i \pi T}{N}(2k-1)
\label{eq:mu_k}
\end{equation}

Since the Hamiltonian is symmetric under exchange of particles and
holes if the sign of the chemical potential is changed simultaneously,
we can simplify (\ref{eq:P_v1}) to
\begin{equation}
P_{N,m}(\mu(j))=
\frac{2 i}{N}\sum_{k=1}^{\lfloor N/2 \rfloor}
\sin\left(\pi m\frac{2k-1}{N}\right)\delta(\mu(j)-\mu_k)
\label{dfu}
\end{equation}
where $\lfloor N/2 \rfloor$ denotes the integer part of $N/2$.  As the
discussion below will show, this is the minimal representation of the
distribution function corresponding to the minimal set of the discrete
imaginary chemical potentials. Another distributions function
different from (\ref{dfu}) can be constructed when the sum 
is taken from $k=N/2+1$ to $N$. Nevertheless, this DF is
different from (\ref{dfu}) only by the sign of imaginary chemical
potentials $\tilde \mu_k=\mu_k^*=-\mu_k$ and thus is supplementary to (\ref{dfu}).

Particularly interesting for even $N$ is the case when the SU($N$)
orbital is half--filled, $m=N/2$.  Then all chemical potentials are
weighted with equal weight
\begin{equation}
P_{N,N/2}(\mu(j))=\frac{2i}{N}\sum_{k=1}^{N/2}(-1)^{k+1}\delta(\mu(j)-\mu_k)
\label{nn2}
\end{equation}
Taking the limit $N\to\infty$ one may replace the summation in
expression (\ref{nn2}) in a suitable way by integration.  Note, that
taking $N\to\infty$ and $m\to\infty$ we nevertheless keep the ratio
$m/N=1/2$ fixed. Then, the
following limiting distribution function
can be obtained:
\begin{equation}
P_{N,N/2}(\mu(j)) \stackrel{N\to\infty}{\longrightarrow}
\frac{\beta}{2\pi i}
\exp\left(-\beta \mu(j) \frac{N}{2}\right)
\label{pg}
\end{equation}
resulting in the usual continuous representation of the local
constraint for the simplest case $n_c=1$ (compare it with (\ref{part1}))
\begin{equation}
Z_S=Tr(\exp\left(-\beta H_F\right) \delta (n_j -\frac{N}{2}))
\end{equation}
We note the obvious similarity of the limiting DF (\ref{pg}) with the
{\it Gibbs canonical distribution} provided that 
the Wick rotation from the imaginary axis of the chemical potential $\mu$
to the real axis of energies $E$ is performed and thus $\mu(j) N/2$ has a
meaning of energy.

Up to now the representation we discussed was purely fermionic and
expressed in terms of usual Grassmann variables when the path integral
formalism is applied. The only difference from slave fermionic
approach is that imaginary chemical potentials are introduced to
fulfill the constraint.  Nevertheless, by making the replacement
$$  
a_k(j,\tau)) \to a_k(j,\tau)
  \exp\left(\frac{i\pi\tau}{\beta} \frac{2k-1}{N}\right)
$$
\begin{equation}
  \bar a_k(j,\tau) \to \bar a_k(j,\tau)
  \exp\left(-\frac{i\pi\tau}{\beta} \frac{2k-1}{N}\right)
\end{equation}
we are coming to generalized Grassmann (semi-fermionic) boundary
conditions
$$
 a_k(j,\beta) = a_k(j,0)\exp\left(i\pi \frac{2k-1}{N}\right)
$$
\begin{equation}
          \bar a_k(j,\beta) = \bar a_k(j,0)\exp\left(-i\pi
            \frac{2k-1}{N}\right)
\label{bk}
\end{equation}
This leads to a temperature diagram technique for Green functions
\begin{equation}
{\cal G}^{\alpha\beta}(j,\tau)=-
\langle T_\tau a_\alpha(j,\tau) \bar a_\beta(j,0)\rangle
\label{itf}
\end{equation}
of semi-fermions with Matsubara frequencies different from both Fermi
and Bose representations.

\begin{figure}[h]
\begin{center}
  \epsfxsize8cm \epsfysize5cm \epsfbox{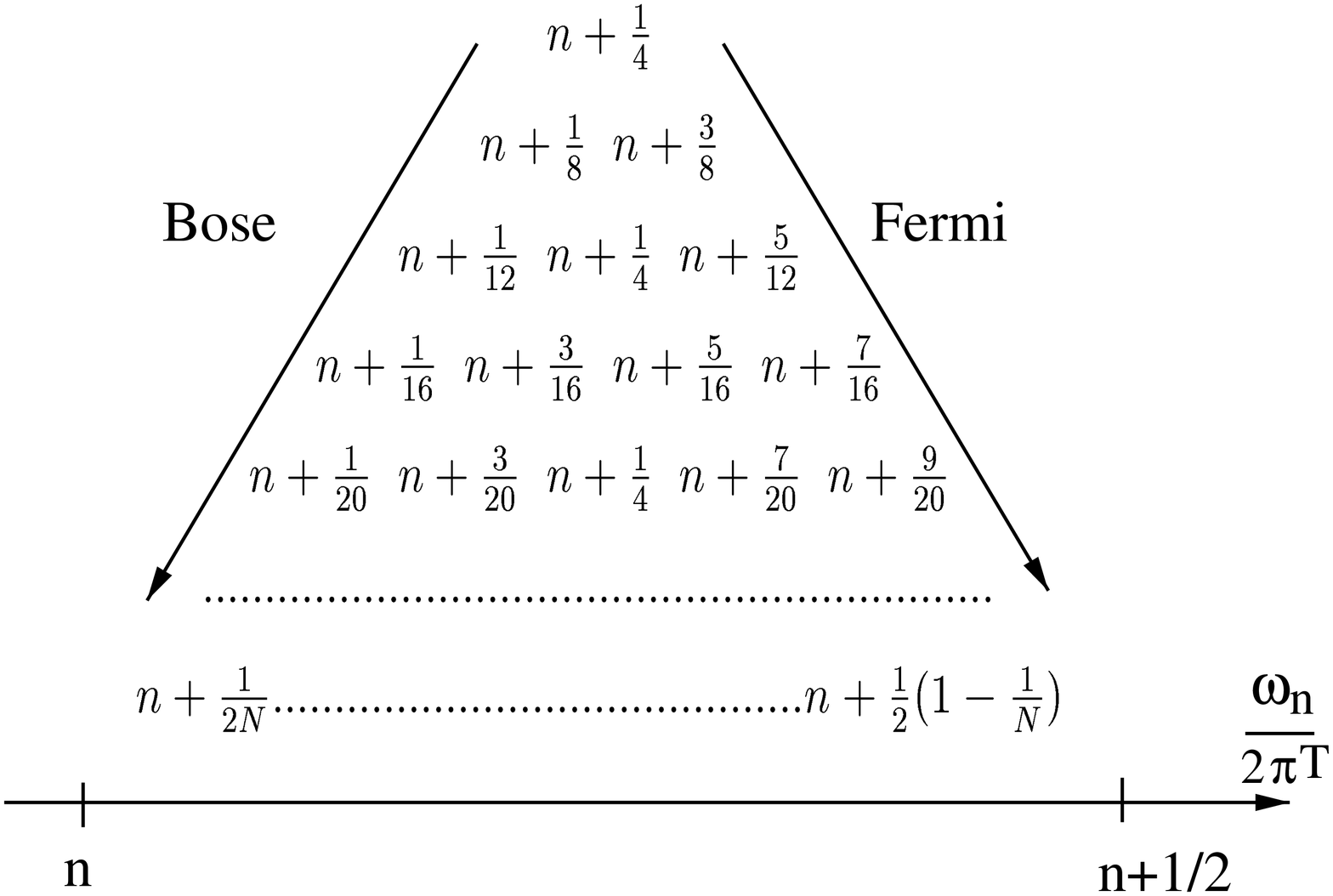}
\end{center}
\caption{The minimal set of Matsubara frequencies for $SU(N)$ 
  representation with even $N$.}
\label{fig:frequencies_N}
\end{figure}

The minimal set of Matsubara frequencies $\omega_n/(2\pi T)$ forms for
SU($N$) with even $N$ the triangle table shown in
Fig.\ref{fig:frequencies_N}.

\begin{figure}
\begin{center}
  \epsfxsize8cm \epsfbox{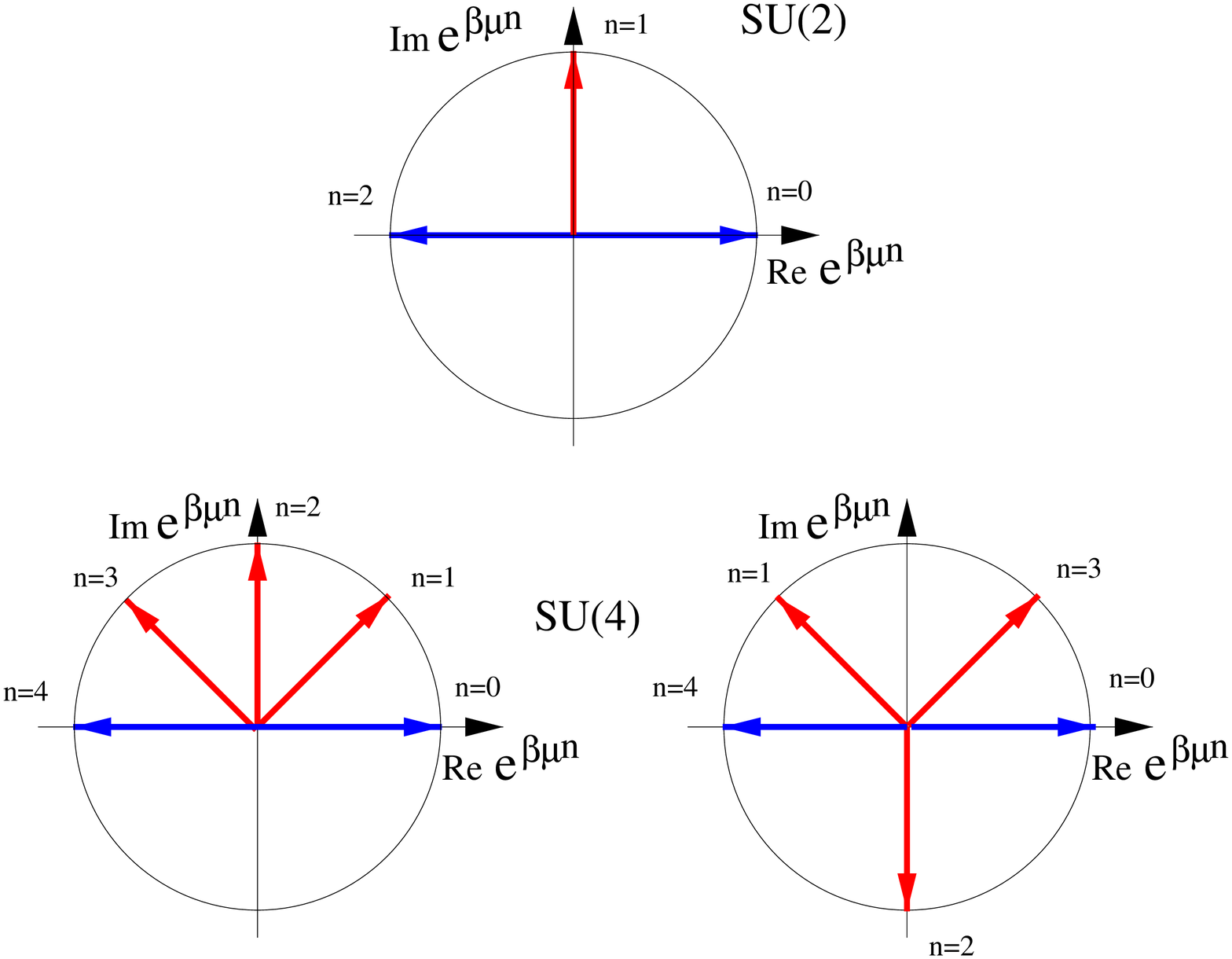}
\end{center}
\caption{Graphical representation of exclusion principle for SU($N$) 
  semi-fermionic representation with even $N$, $n_c=1$ 
(we use $\mu=i\pi T/2$ for SU(2) and
$\mu_1=i\pi T/4,\;\mu_2=3 i \pi T/4$ for SU(4)).}
\label{fig:SU(N)_rep}
\end{figure}

The exclusion principle for this case is illustrated on
Fig.\ref{fig:SU(N)_rep}, where the first two groups SU(2) and SU(4)
are shown. The first point to observe is that the spin Hamiltonian
does not distinguish the $n$ particle and the $n$ hole (or $N-n$
particle) subspace. Due to Eq. (\ref{eq:mu_k}) the two phase factors
$\exp(\beta \mu n)$ and $\exp(\beta \mu (N-n))$ accompanying these
subspaces in Eq. (\ref{dfu}) add up to a purely imaginary value within the
same chemical potential, and the empty and the fully occupied states are
always canceled. In the case of $N \geq 4$, where we have multiple
chemical potentials, the distribution function $P(\mu)$ linearly
combines these imaginary prefactors to select out the desired physical
subspace with particle number $n=m$.

In Fig.\ref{fig:SU(N)_rep}, we note that on each picture the empty and
fully occupied states are canceled in their own unit circle. For SU(2)
there is a unique chemical potential $\mu=\pm i\pi T/2$ which results in
the survival of single occupied states. For SU(4) there are two chemical
potentials (see also Fig.\ref{fig:frequencies_N}). 
The cancellation of single and triple
occupied states is achieved with the help of proper weights for these
states in the distribution function whereas the states with the
occupation number 2 are doubled according to the expression
(\ref{nn2}). In general, for SU($N$) group with $n_c=1$ there exists
$N/2$ circles providing the realization of the exclusion principle.

We consider now the generalization of the SU(2) algebra for the case
of a large moment $S$ with $2S+1$ projections. 
Here, the most convenient fermionic representation is constructed with
the help of a $2S+1$ component Fermi field $a_k(j)$ provided that the
generators of SU(2) satisfy the following equations
$$
  S^+=\sum_{k=-S}^{S-1}\sqrt{S(S+1)-k(k+1)}a^\dagger_{k+1}(j)a_k(j)
$$
$$
  S^-=\sum_{k=-S+1}^{S}\sqrt{S(S+1)-k(k-1)}a^\dagger_{k-1}(j)a_k(j)
$$
\begin{equation}
  S^z=\sum_{k=-S}^S k a^\dagger_{k}(j)a_k(j)
\label{su2}
\end{equation}
such that $dim H_F=2^{2S+1}$ whereas the constraint reads as follows
\begin{equation}
n_j=\sum_{k=-S}^{k=S}a^\dagger_{k}(j) a_{k}(j)=l=1
\label{const2}
\end{equation}
We consider the distribution function for arbitrary $l$ for the sake
of generality. It describes the orbital part of an atomic subshell
with orbital quantum number $S$, with $l$ particles present.  We
denote the corresponding distribution function of the chemical
potential by $P_{2S+1,l}(\mu(j))$.  Following the same routine as for
SU($N$) generators and using the occupancy condition to have $l$ (or
$2S+1-l$) states from the $2S+1$ states filled, one gets the following
distribution function, after using the particle--hole symmetry of
$H_S$:
\begin{equation}
P_{2S+1,l}(\mu(j))=\frac{2i}{2S+1}\sum_{k=1}^{\lfloor S+1/2\rfloor}
\sin\left(\pi l\frac{2k-1}{2S+1}\right)\delta(\mu(j)-\mu_k)
\label{lsu}
\end{equation}
where the chemical potentials are $\mu_k=-i\pi T(2k-1)/(2S+1)$ and
$k=1,...,\lfloor S+1/2 \rfloor$, similarly to Eq.(\ref{eq:mu_k}).

In the particular case of the SU(2) model with $l=1$ for some chosen
values of spin $S$ the distribution functions are determined by the
following expressions
\begin{equation}
  P_{2,1}(\mu(j))=i\;\delta(\mu(j)+\frac{i\pi T}{2})
\nonumber
\end{equation}
for $S=1/2$
\begin{equation}
 P_{3,1}(\mu(j))=P_{3,2}(\mu(j))=
  \frac{i}{\sqrt{3}}\;\delta(\mu(j)+\frac{i\pi T}{3})
\nonumber
\end{equation}
for $S=1$.

This result corresponds to the original Popov-Fedotov description
restricted to the $S=1/2$ and $S=1$ cases. We present as an example
some other distribution functions obtained according to general scheme
considered above:
$$
 P_{4,1}(\mu)=P_{4,3}(\mu)=
$$
\begin{equation}
 =\frac{i\sqrt{2}}{4}\left(
    \delta(\mu+\frac{i\pi T}{4})+
    \delta(\mu+\frac{3i\pi T}{4}\right)
\nonumber
\end{equation}
for $S=3/2$, SU(2) and
\begin{equation}
  P_{4,2}(\mu)=\frac{i}{2}\left(\delta(\mu+\frac{i\pi T}{4})-
    \delta(\mu+\frac{3i\pi T}{4})\right)
\nonumber
\end{equation}
for effective spin "$S=1/2$", SU(4),

$$
 P_{5,1}(\mu)=P_{5,4}(\mu)=
$$
\begin{equation}\displaystyle
 =\frac{i}{\sqrt{10}}\left(\sqrt{1-\frac{1}{\sqrt{5}}}
    \delta(\mu+\frac{i\pi T}{5})+
   \sqrt{1+\frac{1}{\sqrt{5}}} \delta(\mu+\frac{3i\pi T}{5})\right)
\nonumber
\end{equation}
for $S=2$, $SU(2)$ etc.

A limiting distribution function corresponding to Eq. (\ref{pg}) for
the constraint condition with arbitrary $l$ is given by
\begin{equation}
P_{\infty,l}(\mu(j))  \stackrel{S\to\infty}{\longrightarrow}
\frac{\beta}{2\pi i}\exp(-\beta 
l\mu(j))
\end{equation}
For the  case $l=m=N/2\to\infty$ and
$S=(N-1)/2\to\infty$  
the expression for the limiting DF $P_{\infty,l}(\mu(j))$ coincides with (23).
We note that in $S\to\infty$ (or $N\to\infty$) limit continuum
chemical potentials play role of additional U(1) fluctuating field
whereas for finite $S$ and $N$ they are characterized by fixed and 
discrete values.

\begin{figure}
\begin{center}
  \epsfxsize8cm \epsfysize5cm \epsfbox{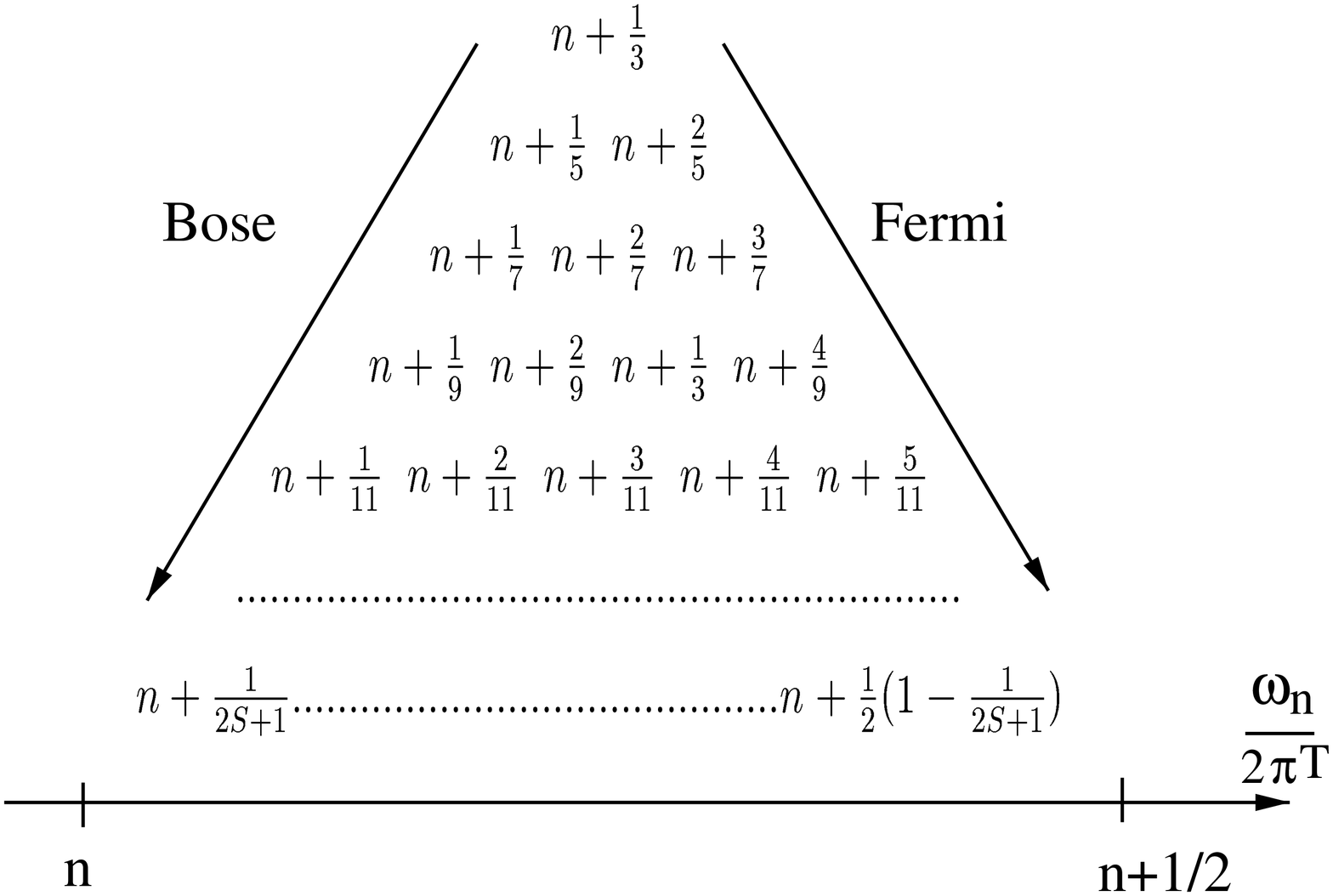}
\end{center}
\caption{The minimal set of Matsubara frequencies for $SU(2)$ 
  representation for integer values of the spin and $l=1$.}
\label{fig:frequencies_spin}
\end{figure}

When $S$ assumes integer values, the minimal fundamental set of
Matsubara frequencies is given by the table in
Fig.\ref{fig:frequencies_spin}.

\begin{figure}
\begin{center}
  \epsfxsize9cm \epsfbox{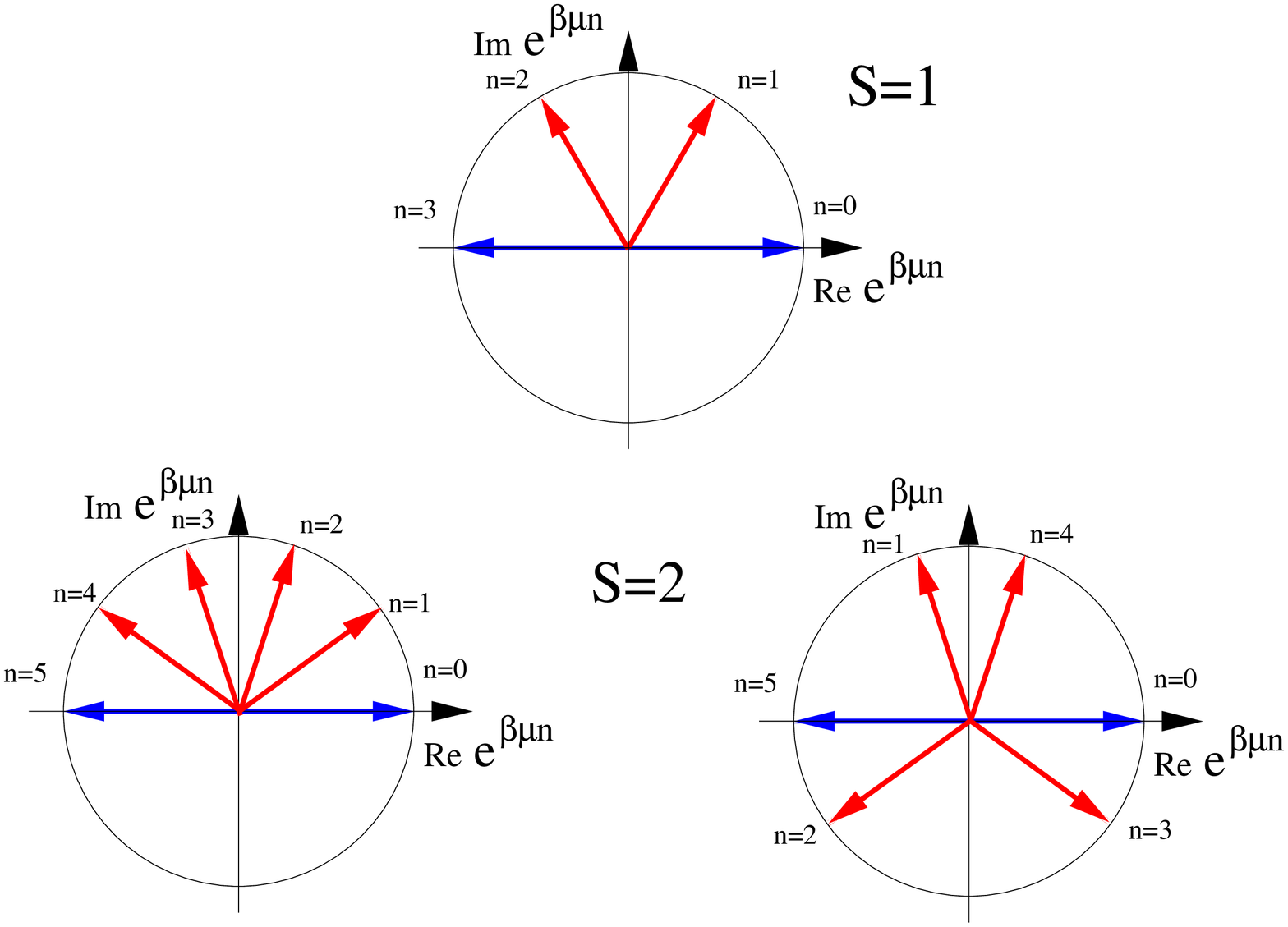}
\end{center}
\caption{Graphical representation of exclusion principle for SU(2) 
  semi-fermionic representation for $S=1$  and $S=2$. 
  For any arbitrary integer
  value of spin there exists $S$ circle diagrams corresponding to the
  $S$ different chemical potentials and providing the realization of
  the exclusion principle.}
\label{fig:spin_rep}
\end{figure}

The exclusion principle for SU(2) in the large spin limit can be also
understood with the help of Fig.\ref{fig:SU(N)_rep} and 
Fig.\ref{fig:spin_rep}.  One can see that empty and fully occupied states
are canceled in each given circle similarly to even-$N$ SU($N$)
algebra.  The particle-hole (PH) symmetry of the representation results in
an equivalence of single occupied and $2S$ occupied states whereas all
the other states are canceled due to proper weights in the
distribution function (\ref{lsu}). In accordance with PH symmetry
being preserved for each value of the chemical potential all circle diagrams
(see Fig.3, Fig.5) are invariant with respect to simultaneous change
$\mu \leftrightarrow -\mu$ and $n_{particle} \leftrightarrow n_{holes}$. 

Let us make few comments concerning the general rectangular Young
tableau of size $n_c * m$. The fermionic representation (\ref{fermi})
is characterized by an $N * n_c$ component field with $n_c$ identical
diagonal constraints and $n_c(n_c-1)/2$ offdiagonal constraints
(\ref{gconst}). The effective "filling" determining the number
of fermions on each site is $m n_c$.  However, not all of these
$ (n_c N)!/((n_c m)!((N-m)n_c)!)$ states are representing proper
physical states.  One should take into account the constraints 
Eq.(\ref{gconst}) to obtain the complete set. 
The number of physical states of a rectangular Young tableau is 
given by the expression:
$$
\nu(N,m,n_c)=\frac{\overbrace{C_{N+n_c-1}^m C_{N+n_c-2}^m...C_{m+n_c}^m}^{N-m}}
{\underbrace{C_{N-1}^m C_{N-2}^m...C_{m}^m}_{N-m}}=
$$
\begin{equation}
=
\frac{\overbrace{C_{N+n_c-1}^m C_{N+n_c-2}^m...C_{N}^m}^{n_c}}
{\underbrace{C_{m+n_c-1}^m C_{m+n_c-2}^m...C_{m}^m}_{n_c}}
\end{equation}
While the diagonal part of Eq.(\ref{gconst}) could be satisfied with
color--dependent chemical potentials $\mu_{a}(j)$, coupling only to
$\sum_\alpha c^\dagger_{\alpha a} c^{\alpha a}$, an exclusion procedure
for the off--diagonal constraints needs either projection operators
or an a priori restriction on the trace, using e.g. coherent states\cite{sachdev89a}.

Another generalization is applicable for a broader range of cases.
The general Young tableau (not necessarily rectangular), representing any
 irreducible representation $(p)$,
can be described in the context of our approach in the following way.
The generators $S_m^{(p)}$ are expressed as matrices
\begin{equation}
(S_m^{(p)})_{\beta \alpha}
=\langle \psi_\beta^{(p)} | T_m^{(k) \times (l)} | \psi_\alpha^{(p)} \rangle
\end{equation}
with $T_m^{(k) \times (l)}= T_m^{(k)}+T_m^{(l)}$ being the generators in a 
suitable direct product of representations $(k)$ and $(l)$ and the states 
$|\psi_\alpha^{(p)}\rangle$ are obtained in terms of Clebsch--Gordon coefficients
\begin{equation}
|\psi_\alpha^{(p)}\rangle = \sum_{\mu,\nu} (k,l;\mu,\nu|p;\alpha)
|\psi_\mu^{(k)}\rangle \times |\psi_\nu^{(l)}\rangle
\end{equation}
This procedure can be iterated until $(k)$ and $(l)$ are fundamental irreducible representations of SU($N$).
The size of the matrices $S_m^{(p)}$ is equal to the dimension of the representation, $\nu^{(p)}$. The trace is now easily expressed in terms of $\nu^{(p)}$
fermionic fields, enforcing the constraint $\delta_{n_j,1}$ with the 
distribution of chemical potentials (see Eq.(\ref{dfu}))
\begin{equation}
P_{\nu^{(p)},1}(\mu(j))=
\frac{2 i}{\nu^{(p)}}\sum_{k=1}^{\lfloor \nu^{(p)}/2 \rfloor}
\sin\left(\pi \frac{2k-1}{\nu^{(p)}}\right)\delta(\mu(j)-\mu_k)
\end{equation}
For the simple case of SU(2), which yields only single--row tableaux, 
this procedure gives the fermionic representation described in 
Eqs.(\ref{su2}-\ref{lsu}). In the case of single--column
tableaux for SU($N$), however, and in the general case of mixed symmetry,
it does not fully use the fermionic commutation properties. Therefore, it
is in general not the representation with the minimal number of fermions
and the minimal number of chemical potentials in $P(\mu)$.

We discuss finally the real-time formalism based on the semi-fermionic
representation of SU($N$) generators. This approach is necessary for
treating the systems being out of equilibrium, especially for many
component systems describing Fermi (Bose) quasiparticles interacting
with spins. The real time formalism is also an alternative approach
for the analytical continuation method for equilibrium problems
allowing direct calculations of correlators whose analytical
properties as function of many complex arguments can be quite
cumbersome.

A long time ago Keldysh \cite{keldysh} and Schwinger \cite{schwinger}
have proposed a novel approach
for the description of kinetic phenomena in metals. This approach
was found especially fruitful for normal metals \cite{smith}, and, in many
recent applications, for superconductors \cite{larkin}, for
disordered interacting (normal or superconducting) electron liquids
\cite{kamenev1} for example. The previous application of the real-time
formalism to the quantum theory of Bose-Einstein condensation (BEC)
\cite{stoof}
allowed the derivation of a Fokker-Planck equation, which describes both
kinetic and coherent stages of BEC. Moreover \cite{lozano} developed the
closed-time path integral formalism for aging effects in quantum
disordered systems being in contact with an environment.
The  Keldysh technique in application to disordered systems (see 
\cite{kamenev1}-\cite{kamenev2} and \cite{kree}
-\cite{sompol}) has also been recently applied to develop a field theory 
alternative to the previously used  replica technique.

\begin{figure}
\begin{center}
  \epsfxsize8cm \epsfbox{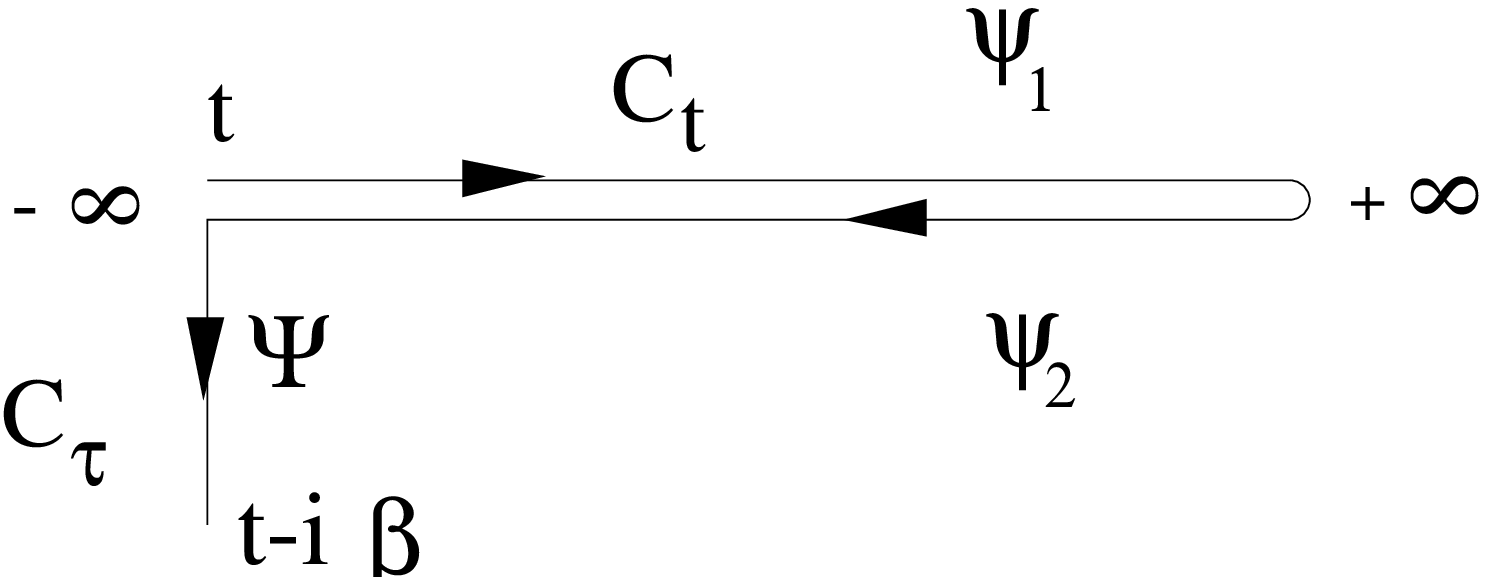} \vspace*{5mm}
\end{center}
\caption{The Keldysh contour going from $-\infty\to\infty\to-\infty$ in real 
  time. The boundary conditions on the imaginary time segment
  determine the generalized distribution functions for
  quasiparticles.}
\label{fig:keldysh}
\end{figure}

To derive the real-time formalism for SU($N$) generators we use the
path integral representation along the closed time Keldysh contour
(see Fig.\ref{fig:keldysh}).
Following the standard route \cite{babichenko86a} we can express the
partition function of the problem containing SU($N$) generators as a
path integral over Grassmann variables
$\psi_l=(a_{l,1}(j),...,a_{l,N}(j))^{T}$ where $l=1,2$ stands for
upper and lower parts of the Keldysh contour, respectively,
\begin{equation}
{\cal Z}/{\cal Z}_0=\int D\bar\psi D\psi\exp(i{\cal A})/
\int D\bar\psi D\psi\exp(i{\cal A}_0)
\label{pfunk}
\end{equation}
where the actions ${\cal A}$ and ${\cal A}_0$ are taken as an integral
along the closed-time contour $C_t+C_\tau$ which is shown in Fig.\ref{fig:keldysh}.
The contour is closed at $t=-\infty+i\tau$ since $\exp(-\beta
H_0)=T_\tau\exp\left(-\int_0^\beta H_0 d\tau\right).$ We denote the
$\psi$ fields on upper and lower sides of the contour $C_t$ as
$\psi_1$ and $\psi_2$ respectively. The fields $\Psi$ stand for the
contour $C_\tau$. These fields provide matching conditions for
$\psi_{1,2}$ and are excluded from final expressions.  Taking into
account the semi-fermionic boundary conditions for generalized
Grassmann fields (\ref{bk}) one gets the matching conditions for
$\psi_{1,2}$ at $t=\pm\infty$,
$$
\psi^\mu_{1,\alpha}|_k(-\infty) =
  \exp\left(i\pi\frac{2k-1}{N}\right)\psi^\mu_{2,\alpha}|_k(-\infty)
$$
\begin{equation}
  \psi^\mu_{1,\alpha}|_k(+\infty) =\psi^\mu_{2,\alpha}|_k(+\infty)
\end{equation}
for $k=1,...,\lfloor N/2 \rfloor$ and $\alpha=1,...,N$.  The
correlation functions can be represented as functional derivatives of
the generating functional
\begin{equation}
Z[\eta]={\cal Z}_0^{-1}\int D\bar\psi D\psi\exp\left(i{\cal A}+
i\oint_C d t (\bar\eta\sigma^z\psi+\bar\psi\sigma^z \eta)\right)
\end{equation}
where $\eta$ represents sources and the $\sigma^z$ matrix stands for
"causal" and "anti-causal" orderings along the contour.

The on-site Green's functions (GF) which are matrices of size
$2N\times 2N$ with respect to both Keldysh (lower) and spin-color
(upper) indices are given by
\begin{equation}
G_{\mu\nu}^{\alpha\beta}(t,t')=
-i\frac{\delta}{i\delta\bar \eta_\mu^\alpha(t)}
\frac{\delta}{i\delta \eta_\nu^\beta(t')}
Z[\eta]|_{\bar\eta,\eta\to 0}
\label{rtf}
\end{equation}
To distinguish between imaginary-time (\ref{itf}) and 
real-time (\ref{rtf}) GF's
we use different notations for Green's functions in these representations.

After a standard shift-transformation \cite{babichenko86a} of fields $\psi$
the Keldysh GF of free semi-fermions assumes the form
\begin{eqnarray}
G_0^\alpha(\epsilon)=G^{R,\alpha}_0
\left(
\begin{array}{cc}
1 - f_\epsilon &  -f_\epsilon\\
1 - f_\epsilon &  -f_\epsilon
\end{array}\right)-
G^{A,\alpha}_0
\left(
\begin{array}{cc}
-f_\epsilon & -f_\epsilon\\
1 - f_\epsilon & 1 - f_\epsilon
\end{array}
\right)
\nonumber
\end{eqnarray}
where the retarded and advanced GF's are
\begin{equation}
G^{(R,A)\alpha}_0(\epsilon)=(\epsilon \pm i\delta)^{-1},
\quad
f_\epsilon=f^{(N,k)}(\epsilon)
\end{equation}
with equilibrium distribution functions
\begin{equation}
f^{(N,k)}(\epsilon)=T\sum_n\frac{e^{i\omega_{n_k}\tau|_{+0}}}
{i\omega_{n_k}-\epsilon}=
\frac{1}{e^{i\pi (2k-1)/N}\exp(\beta\epsilon)+ 1}
\end{equation}
A straightforward calculation of $f^{(N,k)}$ for the case of even $N$
leads to the following expression
$$
f^{(N,k)}(\epsilon)=
$$
\begin{equation}
=\frac{\displaystyle\sum_{l=1}^N(-1)^{l-1}
\exp\left(\beta\epsilon(N-l)\right)
\exp\left(-\frac{i\pi l(2k-1)}{N}\right)}{\exp(N\beta\epsilon)+1},
\end{equation}
where $k=1,...,N/2.$

The equilibrium distribution functions (EDF) $f^{(2S+1,k)}$ for the
auxiliary Fermi-fields representing arbitrary $S$ for $SU(2)$ algebra
are given by 
$$
f^{(2S+1,k)}(\epsilon)=
$$
\begin{equation}
=\frac{\displaystyle\sum_{l=1}^{2S+1}(-1)^{l-1}
\exp\left(\beta\epsilon(2S+1-l)\right)
\exp\left(-\frac{i\pi(2k-1)}{2S+1})\right)}
{\exp((2S+1)\beta\epsilon)+(-1)^{2S+1}}
\end{equation}
for $k=1,...,\lfloor S+1/2 \rfloor$.  Particularly simple are the
cases of $S=1/2$ and $S=1$,
$$
 f^{(2,1)}(\epsilon)=
    n_F(2\epsilon)-i\frac{1}{2\cosh(\beta\epsilon)}
$$
\begin{equation}
     f^{(3,1)}(\epsilon)=
    \frac{1}{2}n_B(\epsilon)-\frac{3}{2}n_B(3\epsilon)
    -i\sqrt{3}\frac{\sinh(\beta\epsilon/2)}{\sinh(3\beta\epsilon/2)}
\end{equation}
Here, standard notations for Fermi/Bose distribution functions
$n_{F/B}(\epsilon)=[\exp(\beta\epsilon) \pm 1]^{-1}$ are used.  

In general the EDF for half-integer and integer spins can be expressed in
terms of Fermi and Bose EDF respectively.  We note that since
auxiliary Fermi fields introduced for the representation of SU($N$)
generators do not represent the true quasiparticles of the problem,
helping only to treat properly the constraint condition, the
distribution functions for these objects in general do not have to be
real functions.  Nevertheless, one can prove that the imaginary part
of the EDF does not affect the physical correlators and can be
eliminated by introducing an infinitesimally small real part for the
chemical potential.  In spin problems, a uniform/staggered magnetic
field usually plays the role of such real chemical potential for 
semi-fermions.

Let us illustrate the application of the semi-fermionic formalism for spin
Hamiltonians. As an example we consider the SU(2)
Heisenberg model for $S=1/2$ with the nearest neighbor
interaction
\begin{equation}
H_{int}=-\sum_{<ij>}J_{ij}\vec{S}_i\vec{S}_j
\label{hh1}
\end{equation}
  
We start with imaginary-time semi-fermionic description of the ferromagnetic 
(FM) state of the Heisenberg model ($J > 0$).
We follow the standard procedure
developed in the original paper of Popov and Fedotov \cite{popov88a}. 
After applying the Hubbard-Stratonovich transformation to decouple the 
four-semi-fermion term in (\ref{hh1}) by 
the {\it local vector} field $\vec{\Phi}_i(\tau)$ 
the effective action is obtained:

$${\cal A}^{FM}_{eff}[\psi,\Phi]=\tilde{\cal
A}_0[\psi,\Phi]-$$
\begin{equation}
- \frac{1}{4}\int_0^\beta d\tau
\sum_{\vec{q}}(I_{FM}(\vec{q}))^{-1}\vec{\Phi}_q(\tau)\vec{\Phi}_q(\tau)
\end{equation}
where $\psi^T=(\psi_\uparrow \psi_\downarrow)$ - 
fields denoting the semi-fermions in SU(2) representation of the $S=1/2$ 
spin operators,
\begin{equation}
I_M(q)=I_{FM/AFM}(q)=\frac{1}{N}\sum_{r_{ij}}
J (\vec{r}_{ij})e^{i\vec{q}\vec{r}}
\end{equation}
and $I_M(0)=ZJ>0$ for the FM instability (here $Z$ denotes the number of the
nearest neighbors, N stands for the number of unit cells). 
The FM phase transition corresponds to appearance at $T=T_c$ 
of the nonzero average $\langle\Phi^z(0,0)\rangle$ which stands 
for nonzero uniform magnetization, or by another words, 
corresponds to the Bose condensation of the field $\Phi^z$.

Splitting the field $\Phi^z$ on the 
time-independent spatially homogeneous ({\it uniform}) 
part and the fluctuating filed
$\tilde\Phi^z$
\begin{equation}
\Phi^z(\vec{k},\omega)={\cal
M}(\beta N)^{1/2}\delta_{\vec{k},0} \delta_{\omega,0}
+\tilde\Phi^z(\vec{k},\omega)
\end{equation} 
make it possible to integrate over all semi-fermionic fields.
As a result, the nonpolynomial
effective action can be derived for the FM Heisenberg model 
\begin{equation}
{\cal
A}_{eff}={\cal A}_0[\Phi]+ {\tt Tr} \ln\left({\cal
G}_\sigma^{-1}(\Phi^z,\Phi^\pm)\right)
\end{equation}
where ${\cal G}_\sigma=-\langle T_\tau \psi_\sigma(\vec{j},\tau)
\bar\psi_\sigma(\vec{j},0\rangle$ stands for the 
local Green's function of semi-fermions.
The expansion of the $Tr\ln {\cal G}_\sigma^{-1}$ with respect to $\Phi$ fields
results in standard Ginzburg-Landau functional (see Fig.7). 
The effects of molecular 
field are included into zero approximation for GF:
$${\cal G}^{0}_{\sigma}(i\omega_n)=[i\omega_n +
\sigma^z_{\sigma\sigma}{\cal M}/2]^{-1}.$$
\begin{figure}
\begin{center}
\epsfxsize6cm
\epsfbox{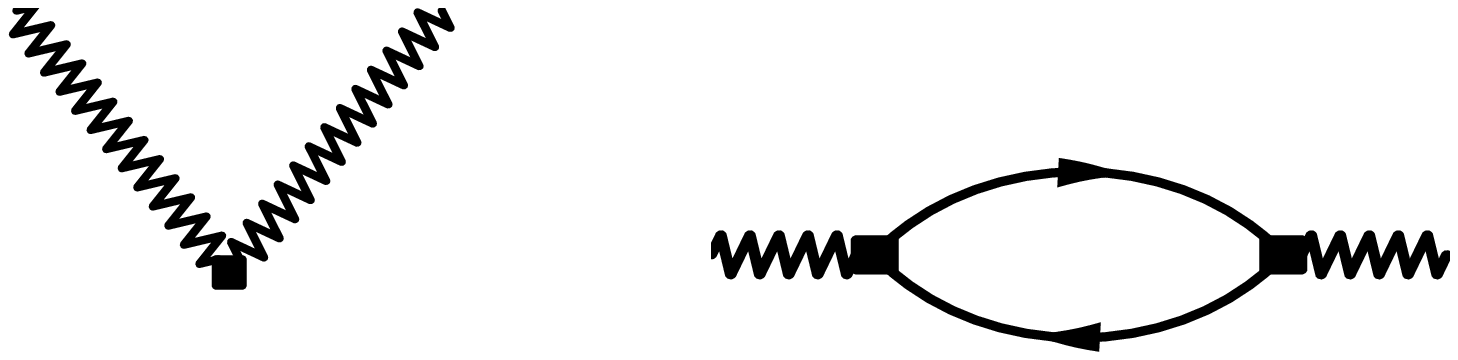}
\caption{First few graphs for the free energy expanded 
with respect to local molecular magnetic field. Solid line stands 
for semi-fermions. Zig-zag line denotes the "condensate" field.}
\end{center}
\label{freee2}
\end{figure}

In one loop approximation the
standard molecular field equation can be reproduced 
\begin{equation}
{\cal M}=I_{FM}(0)\tanh(\beta{\cal M}/2)
\label{sucun}
\end{equation}

The saddle point (mean-field) effective action is given by 
well-known expression
\begin{equation}
{\cal A}_0[{\cal M}]=-N\left[\frac{\beta {\cal M}^2}{4I_M(0)}-
\ln\left(2\cosh\left(\frac{\beta {\cal M}}{2}\right)\right)\right]
\end{equation}
and the free energy per spin $f_0$ (see Fig.7) is determined by standard
equation:
\begin{equation}
\beta f_0 =-\ln Z_S=\frac{\beta {\cal M}^2}{4I_M(0)}-
\ln\left(2\cosh\left(\frac{\beta {\cal M}}{2}\right)\right)
\end{equation}

Calculation of the second variation of ${\cal A}_{eff}$ gives rise to
the following expression
$$\delta{\cal A}_{eff}=-\frac{1}{4}\sum_{\vec{k}}\Phi^z(\vec{k},0)
\left[I_M^{-1}(\vec{k})-\frac{\beta}{2\cosh^2(\beta\Omega)}\right]
\Phi^z(\vec{k},0)-$$
$$-\frac{1}{4}\sum_{\vec{k},\omega\ne 0}
I_M^{-1}(\vec{k})\Phi^z(\vec{k},\omega)\Phi^z(\vec{k},\omega)-$$
\begin{equation}
-\sum_{\vec{k},\omega}\Phi^+(\vec{k},\omega)\left[
I_M^{-1}(\vec{k})-\frac{\tanh(\beta\Omega)}{2\Omega - i\omega}\right]
\Phi^-(\vec{k},\omega)
\end{equation}
where $\Omega=(g\mu_B H +{\cal M})/2$.  For $T>T_c$ one easily obtains 
the effective static 
spin-spin interaction equivalent to those given by the  
Random Phase Approximation (RPA)
$$\Gamma(\vec{q},0)=
\langle \vec{\Phi}(\vec{q},0)\vec{\Phi}(-\vec{q},0)\rangle=
2 I(\vec{q})/(1-2 \chi_0 I(\vec{q})),$$ 
where $\chi_0^{+-}(\vec{q},0)=2\chi_0^{zz}(\vec{q},0)=2\chi_0=2S(S+1)\beta/3$ 
stands for the on-site spin susceptibility in paramagnetic state.

Let us now consider the Heisenberg model with antiferromagnetic (AFM) sign of
 the exchange integral ($J < 0$).
$${\cal A}^{AFM}_{eff}[\psi,\Phi]=\tilde{\cal A}_0[\psi,\Phi]+$$
\begin{equation}+
\frac{1}{4}\int_0^\beta d\tau
\sum_{\vec{q}}(I_{AFM}(\vec{q}))^{-1}\vec{\Phi}_q(\tau)\vec{\Phi}_q(\tau)
\end{equation}
and $I_M(\vec{Q})=ZJ<0$ for the AFM instability corresponding to
vector ${\bf Q}=(\pi,...,\pi)$ 
(we consider the hypercubic lattice for simplicity).
In contrast to the FM case, we can now represent the longitudinal 
component of the field $\Phi^z$ as a superposition of the  {\it staggered}
time-independent part ("staggered condensate") and a fluctuating field
\begin{equation}
\Phi^z(\vec{k},\omega)={\cal N}(\beta N)^{1/2}\delta_{\vec{k},\vec{Q}} \delta_{\omega,0}
+\tilde\Phi^z(\vec{k},\omega) 
\end{equation}
As a result, the integration over semi-fermionic  
fields can be done explicitely.
Introducing two sublattices for $\psi$ fields 
one gets $4 \times 4$ matrix structure for the semi-fermionic 
Green's functions.
Since the AFM instability is associated with appearance of a
nonzero staggered magnetization ${\cal N}$, 
it is necessary to take into account both "normal"
and "anomalous" GF determined as follows:
$${\cal G}^0_\sigma(i\omega_n)=
-\int_0^\beta d\tau e^{i\omega_n \tau}
\langle T_\tau \psi_\sigma(\vec{k},\tau)\bar \psi_\sigma(\vec{k},0)\rangle=$$
\begin{equation}=
-\frac{i\omega_n}{\omega_n^2+\tilde \Omega^2}
\end{equation}
$${\cal F}^0_\sigma(i\omega_n)=
-\int_0^\beta d\tau e^{i\omega_n \tau}
\langle T_\tau \psi_\sigma(\vec{k},\tau)
\bar \psi_\sigma(\vec{k}+\vec{Q},0)\rangle=$$
\begin{equation}=
-\frac{\tilde \Omega \sigma^z_{\sigma\sigma}}{\omega_n^2+\tilde\Omega^2}.
\end{equation}
where $\tilde \Omega=({\cal N}+g\mu_B h)/2$.
Integrating over all
semi-fermions one obtains the mean-field equation for the
staggered magnetization:
\begin{equation}
{\cal N}=-I_{AFM}(Q)\tanh(\beta{\cal N}/2)
\label{sustag}
\end{equation}
and
\begin{equation}
{\cal A}_0[{\cal N}]=N\left[\frac{\beta {\cal N}^2}{4I_M(Q)}+
\ln\left(2 \cosh\left( \frac{\beta {\cal N}}{2}\right)\right)\right]
\end{equation}

After taking into account the second variation of ${\cal A}_{eff}$ 
the following expression for the effective action is obtained [(see e.g.
\cite{bouis},\cite{azakov}):
$$\delta{\cal A}_{eff}=\frac{1}{4}\sum_{\vec{k}}\Phi^z(\vec{k},0)
\left[I_M^{-1}(\vec{k})+\frac{\beta}{2\cosh^2(\beta\tilde\Omega)}\right]
\Phi^z(\vec{k},0)+$$
$$+\frac{1}{4}\sum_{\vec{k},\omega\ne 0}
I_M^{-1}(\vec{k})\Phi^z(\vec{k},\omega)\Phi^z(\vec{k},\omega)+$$
$$+\sum_{\vec{k},\omega}\Phi^+(\vec{k},\omega)\left[
I_M^{-1}(\vec{k})+\frac{2\tilde\Omega\tanh(\beta\tilde\Omega)}
{4\tilde\Omega^2 +\omega^2}
\right]\Phi^-(\vec{k},\omega)-$$
\begin{equation}
-\sum_{\vec{k},\omega}\Phi^+(\vec{k}+\vec{Q},\omega)
\frac{i\omega}{4\tilde\Omega^2 +\omega^2}
\Phi^-(\vec{k},\omega).
\end{equation}

The application of the Schwinger-Keldysh formalism for the Heisenberg 
model is straightforward. Applying the
semi-fermionic 
transformation to the partition function one obtains the action as an
integral along the closed-time Keldysh-contour
\begin{equation}
{\cal A}={\cal A}_0+{\cal A}_{int}={\cal A}_0+
\oint_{C}d t\sum_{\bf q}J({\bf q})\vec{S}_{\bf q}(t)\vec{S}_{-\bf q}(t)
\label{abeg}
\end{equation}
where ${\cal A}_0$ corresponds to noninteracting semi-fermions
\begin{eqnarray}
{\cal A}_0=\oint_C dt\sum_i\bar\psi_i
\left(
\begin{array}{cc}
(G^{R,\alpha}_0)^{-1} &0\\
0 & (G^{A,\alpha}_0)^{-1}
\end{array}
\right)\psi_i
\label{a0}
\end{eqnarray}
We denote $J_{\bf q}$$=$$J$$\sum_{<{\bf l}>}$$e^{i{\bf ql}}$,
$\nu_{\bf q}$$=$$J_{\bf q}$$/$$J_0$
and apply  four - component semi-fermionic  representation  for FM case and 
eight - component representstion with
$\psi^T$$=$$(\tilde\psi^T_{\bf k}$$\tilde\psi^T_{\bf k+Q})$ for the AFM case.
Performing the standard Hub\-bard - Stra\-to\-no\-vich 
transformation along the Keldysh 
contour with the help of the two-Keldysh-component {\it vector} (Bose)
field $\Phi$, one gets
\begin{equation}
{\cal A}_{int}=-\frac{1}{2}Tr(\vec{\Phi}_{\bf q}^T
J^{-1}_{\bf q}\sigma^z \vec{\Phi}_{\bf q})
+Tr(\bar\psi\vec{\Phi}_\mu\vec{\sigma}\gamma^\mu\psi)
\label{aint}
\end{equation}
Now we integrate out $\psi$ fields and express the effective
action in terms of $\vec\Phi$ fields
$$ {\cal A}_{eff}=
-\frac{1}{2}Tr(\vec{\Phi}_{\bf q}^T J^{-1}_{\bf q}\sigma^z \vec{\Phi}_{\bf q})
+Tr\ln\left(G_0^{-1}+\vec{\Phi}_\mu\vec{\sigma}\gamma^\mu\right)
$$
where $\gamma^\mu$$=$$(\sigma^z\pm 1)/2$ acts in Keldysh space.
Since in general $\vec\Phi$ is a time- and space-dependent
fluctuating field the partition function (\ref{pfunk}) cannot be evaluated
exactly. Nevertheless, when a magnetic instability occurs, we can
represent the longitudinal component of this field as a
superposition of a uniform (FM) or staggered (AFM) time-independent part
and a fluctuating field
\begin{equation}
\Phi^z_\mu({\bf q},\omega)={\it condensate}\;+\;\phi^z_\mu({\bf q},\omega),
\label{cond1}
\end{equation}
where  $\Phi_\mu^{\pm}({\bf q},\omega)=\phi_\mu^\pm({\bf q},\omega)$ with
the matching conditions at $t=\pm\infty$
\begin{equation}
\phi_1^{\alpha}(-\infty)=\phi_2^{\alpha}(-\infty),\;\;
\phi_1^{\alpha}(+\infty)=\phi_2^{\alpha}(+\infty).
\label{bc3}
\end{equation}
We expand
$Tr\ln(G_0^{-1}+\vec{\phi}_\mu\vec{\sigma}\gamma^\mu)$ in accordance with
\begin{equation}
Tr\ln(...)=Tr\ln G_0^{-1}+ Tr\sum_{n=1}^{\infty}
\frac{(-1)^{n+1}}{n}(G_0\vec{\phi}_\mu\vec{\sigma}\gamma^\mu)^n
\label{tr}
\end{equation}

\begin{figure}
\begin{center}
\epsfxsize6cm
\epsfbox{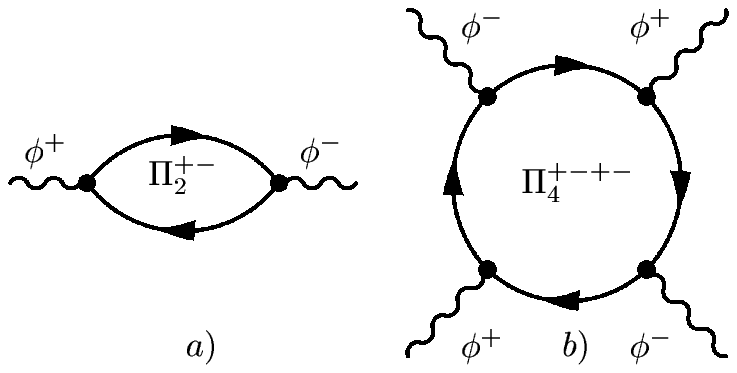}
\caption{Feynman diagrams contributing to dispersion (a) and  damping (b) of
magnons. Solid line denotes semi-fermions.}
\end{center}
\end{figure}

The spectrum of the excitations (FM or AFM magnons) can be defined as poles
of the transverse GF
$$D^{+-}_{{\bf x},t}=D({\bf x},t)=-i
\langle T_C\phi_1^+({\bf x},t)\phi_1^-(0,0)\rangle.$$
The procedure of the calculation of this GF is similar to that for a
"fermionic" GF.
Introducing the sources and evaluating (\ref{tr}) one gets
\begin{eqnarray}
D_0(\omega)=D^{R}_0
\left(
\begin{array}{cc}
1+N_\omega & N_\omega\\
1+N_\omega & N_\omega
\end{array}\right)-
D^{A}_0
\left(
\begin{array}{cc}
N_\omega & N_\omega\\
1+N_\omega  & 1+N_\omega
\end{array}
\right)
\nonumber
\end{eqnarray}
where the retarded and advanced magnons GF's are
$$
D^{R,A}({\bf q},\omega)=(\omega-\omega({\bf q})\pm i\delta)^{-1},\;\;
N_\omega=(\exp(\beta\omega)-1)^{-1}
$$
The magnon spectrum is determined by the zeros of the determinant of
$J^{-1}_{\bf q}-\Pi_2^{+-}(\omega)$ (see Fig.8a) in equilibrium
\begin{equation}
\omega_{\bf q}=J_0{\cal M}(1-\nu_{\bf q})\Rightarrow \lambda {\bf q}^2,
\label{afm2}
\end{equation}
for FM magnons and
\begin{equation}
\omega_{\bf q}=|J_0|{\cal N}\sqrt{1-\nu^2_{\bf q}}\Rightarrow c|{\bf q}|,
\label{afm}
\end{equation}
for AFM magnons. The uniform and staggered magnetization are given 
by equations (\ref{sucun}) and (\ref{sustag}) correspondingly.

The magnon damping is defined by four-magnon processes
$\Pi_4^{+-+-}$, shown in Fig8.b.
The derivation of the kinetic equation and calculation of magnon damping
is reserved here for a detailed publication.

We consider now 
the second possibility to decouple the four-fermion term in the Heisenberg
model with the antiferromagnetic sign of spin-spin interaction which can be 
written in a form equivalent to (\ref{hh1}): 
\begin{equation}
H_{int}=\frac{1}{2}\sum_{<ij>}J_{ij}\psi^\dagger_i\psi_j\psi^\dagger_j\psi_i
+ \frac{1}{4}\sum_{<ij>}J_{ij}\rho_i\rho_j
\label{hh2}
\end{equation}
Last term in the Hamiltonian (\ref{hh2}) describes the fluctuations of
semi-fermionic density 
$\rho_i=\psi^\dagger_i\psi_i$ and therefore is irrelevant
for our calculations.
In contrast to description of the local correlations achieved with the help
of the local {\it vector} bosonic field we introduce now
the bi-local {\it scalar} bosonic field $\Lambda_{ij}$
depending on two sites and responsible for inter-site semi-fermionic 
correlations.
Defining new coordinates $\vec{R}$$=$$(\vec{R}_i$$+$$\vec{R}_j)$$/2$, 
$\vec{r}$$=$$\vec{R}_i$$-$$\vec{R}_j$
and applying a Fourier transformation we obtain the effective action
$$ {\cal A}_{eff}=
-\frac{1}{2}Tr(\Lambda_{Pq_1}^T J^{-1}_{\bf q_1-q_2}\sigma^z \Lambda_{Pq_2})
+Tr\ln\left(G_0^{-1}-\Lambda_\mu\gamma^\mu\right)
$$
This effective action describes the nonequilibrium quantum spin-liquid (SL).
We confine ourselves to consider the uniform phase 
\cite{ioffe},\cite{lee},\cite{rvb1},\cite{rvb2} of
{\it Resonant Valence Bonds} (RVB) in 2D antiferromagnets.
It is suitable to rewrite the functional in new variables, namely
the amplitude $\Delta$  and the phase 
$\Theta$$=$$\vec{r}$$\vec{A}(\vec{R})$,
according to formula
\begin{equation}
\Lambda^{<ij>}_\mu(\vec{R},\;\vec{r})=
\Delta(\vec{r}) J\gamma^\mu\exp\left(i\vec{r}\vec{A}_\mu(\vec{R})\right)
\label{uni}
\end{equation}
The exponent in (\ref{uni}) stands for gauge fluctuations to be taken in
eikonal approximation. As a result,  the effective action can be written  
in continuum limit in terms of the gauge fields $A_\alpha$ as follows:
\begin{equation}
{\cal A}_{eff}=\oint_{C}d t\int d\vec{k}
A_\alpha(\vec{k},t)\pi^{\alpha\beta}A_\beta(\vec{k},t)
\end{equation}
The spectrum of excitations in the uniform SL
is defined by the zeros of current-current correlation function 
\cite{kiselev00b}
$${\Large \pi}^{R,\alpha\beta}_{q,\omega}=Tr(p^\alpha p^\beta
(G^R_{p+q}G^K_p+G^K_{p+q}G^A_p)+\delta_{\alpha\beta}
f(J_{\bf p}\Delta))$$ 
in equilibrium \cite{reizer1,reizer2} 
and is purely diffusive (see e.g.\cite{ioffe})
\begin{equation}
\omega=iJ\Delta |{\bf q}|^3,\;\;\;\;
\Delta=-\sum_{\bf q}\nu({\bf q})\tanh\left(\frac{J_{\bf q}\Delta}{T}\right)
\label{rvb}
\end{equation}
We denote $G^K$ an off-diagonal element (Keldysh component) of semi-fermionic 
GF in triangular representation, provided that
$$G_0^K(\epsilon)=-i2\pi \delta(\epsilon\pm h)[B_{1/2}(\beta\epsilon)\pm i 
{\rm sech}(\beta\epsilon)]$$
is expressed in terms of a Brillouin function $B_{1/2}$ containing 
correct information about occupied states. The equation of motion for $G^K$ 
generally constitutes the quantum kinetic equation.

The quantum kinetic equation for nonequilibrium spin RVB-liquids can be
obtained by taking into account the higher order diagrams
similarly to Fig.8b with current-like vertices and will be presented elsewhere.

We discuss now briefly some possible applications of the imaginary-time and
real-time Schwinger-Keldysh semi-fermionic formalism 
developed for SU(N) Hamiltonians for solution of
the  condensed matter physics problems.
The Keldysh technique in application to disordered systems
attracts  a constant  interest (see \cite{kamenev1}-\cite{lozano})
as an alternative approach to the replica technique.
The main advantage of the closed-time contour calculations
is an automatic normalization (disorder independent) of
the partition function (see \cite{kamenev1}). 
The application of real-time Schwinger-Keldysh approach allows one
to study the quantum dynamics of disordered systems being out of equilibrium.
We note, that the formalism developed in the present
paper is also a very promising tool for description of a 
quantum phase transitions (magnetic, spin-glass etc) in SU(N) models (see 
\cite{sachdev1,sachdev2})
Another possible application of the semi-fermionic SU(N) representation is
the description of paramagnet-(fer\-ro) anti\-fer\-ro\-mag\-net or
paramagnet-spin liquid transitions in equilibrium and non\-equi\-lib\-ri\-um
stron\-gly cor\-re\-la\-ted electron systems (see \cite{kiselev1,kiselev2}). 
The nonlinear spin waves
in strongly correlated lo\-cal - iti\-ne\-rant magnets
and the kinetic properties of the nonequilibrium spin liquid are also 
possible problems to be considered with the method proposed.
The third interesting example of the application of the semi-fermionic formalism is
the Kondo systems \cite{barz},  
for example the Kondo lattice model usually used for interpretation of
an exotic properties of heavy-fermion compounds or the  nonequilibrium 
Kondo-systems in semiconducting hetero-structures (see e.g. \cite{kikoin1}-\cite{kikoin2}). 
The main advantage of the semi-fermionic representation 
in applications to the strongly correlated systems in comparison with another methods is that
the local constraint is taken into account exactly and the usual Feynman
diagrammatic codex for the composite itinerant-local compound is applicable.

Summarizing, we constructed a general scheme for the semi-fermionic
representation for generators of the SU($N$) algebra. A representation
for the partition function is found both in imaginary and real time.
The approach developed leads to the standard diagram technique for
Fermi operators, although the constraint is taken into account
rigorously. The method proposed allows to treat SU($N$) generators on
the same footing as Fermi and Bose systems.  The technique derived can
be helpful for the description of quantum systems in the vicinity of a
quantum phase transition point and for nonequilibrium systems.

This work was started on the workshop "Strongly correlated systems" in
the ICTP in Trieste. One of the authors (MNK) is grateful to many
participants of this workshop and especially N.Andrei, A.Tsvelik and
A.Protogenov for fruitful discussions. We thank also D.Aris\-tov and A.Lu\-ther
for valuable comments and interest to our work. 
The work is supported by the
DFG under SFB 410, by the Evangelisches Studienwerk Villigst (HF) and
by an Alexander von Humboldt fellowship (MNK).

\nocite{haldane91a}


\begin{thebibliography}{99}

\bibitem{holstein40a}
T. Holstein and H. Primakoff, Phys. Rev. B {\bf 58},  1098  (1940).

\bibitem{dyson56a}
F. Dyson, Phys. Rev. {\bf 102},  1217  (1956).

\bibitem{maleev58a}
S.~V. Maleyev, Sov. Phys. JETP {\bf 6},  776  (1958).

\bibitem{abrikosov65a}
A.~A. Abrikosov, Physics {\bf 2},  5  (1965).

\bibitem{abrikosov67a}
A.~A. Abrikosov, Sov. Phys. JETP {\bf 26},  641  (1967).

\bibitem{larkin68a}
V.~G. Vaks, A.~I. Larkin, and S.~A. Pikin, Sov. Phys. JETP {\bf 26},  188
  (1968).

\bibitem{larkin68b}
V.~G. Vaks, A.~I. Larkin, and S.~A. Pikin, Sov. Phys. JETP {\bf 26},  647
  (1968).

\bibitem{hubbard65a}
J. Hubbard, Proc. R. Soc. London A {\bf 285},  542  (1965).

\bibitem{coleman00a}
P. Coleman, C. Pepin, and A.~M. Tsvelik, Phys. Rev. B {\bf 62},  3852  (2000).

\bibitem{coleman00b}
P. Coleman, C. Pepin, and A.~M. Tsvelik, Nuc.Phys {\bf B 586}, 641 (2000) .

\bibitem{coleman00c}
P. Coleman, J. Hopkinson, and C. Pepin, Phys.Rev. {\bf B 63}, 140 411R (2001).

\bibitem{read83a}
N. Read and D.~M. Newns, J. Phys. C {\bf 16},  3273  (1983).

\bibitem{auerbach86a}
A. Auerbach and K. Levin, Phys. Rev. Lett. {\bf 57},  877  (1986).

\bibitem{millis87a}
A.~J. Millis and P.~A. Lee, Phys. Rev. B {\bf 35},  3394  (1987).

\bibitem{bicker87a}
N.~E. Bicker, Rev. Mod. Phys. {\bf 59},  845  (1987).

\bibitem{kotliar86a}
G. Kotliar and A.~E. Ruckenstein, Phys. Rev. Lett. {\bf 57},  1362  (1986).

\bibitem{sachdev89a}
N. Read and S. Sachdev, Nuc. Phys. B {\bf 316},  609  (1989).

\bibitem{sachdev89b}
N. Read and S. Sachdev, Phys. Rev. Lett. {\bf 62},  1694  (1989).

\bibitem{sachdev90a}
N. Read and S. Sachdev, Phys. Rev. B {\bf 42},  4568  (1990).

\bibitem{affleck88a}
I. Affleck and J.~B. Marston, Phys. Rev. B {\bf 37},  3774  (1988).

\bibitem{affleck89a}
J.~B. Marston and I. Affleck, Phys. Rev. B {\bf 39},  11538  (1989).

\bibitem{auerbach88a}
D.~P. Arovas and A. Auerbach, Phys. Rev. B {\bf 38},  316  (1988).

\bibitem{chubukov91a}
A. Chubukov, Phys. Rev. B {\bf 44},  12318  (1991).

\bibitem{gaudin60a}
M. Gaudin, Nuc. Phys. {\bf 15},  89  (1960).

\bibitem{popov88a}
V.~N. Popov and S.~A. Fedotov, Zh. Eksp. Teor. Fiz. {\bf 94},  183  (1988),
  [Sov. Phys. JETP {\bf 67}, 535 (1988)].

\bibitem{oppermann91a}
J. Stein and R. Oppermann, Z. Phys. B {\bf 83},  333  (1991).

\bibitem{gros90a}
C. Gros and M.~D. Johnson, Physica B {\bf 165-166},  985  (1990).

\bibitem{kiselev99a}
F. Bouis and M.~N. Kiselev, Physica B {\bf 259-261},  195  (1999).

\bibitem{kiselev00a}
M.~N. Kiselev and R. Oppermann, JETP Lett. {\bf 71},  250  (2000).

\bibitem{kiselev00b}
M.~N. Kiselev and R. Oppermann, Phys.Rev.Lett {\bf 85},  5631  (2000).

\bibitem{oppermann99b}
H. Feldmann and R. Oppermann, J. Phys. A {\bf 33},  1325  (2000).

\bibitem{cartan}
E. Cartan, {\em Le\c{c}ons sur la theorie des spineurs} (Hermann, Paris, 1938).

\bibitem{auerbach94a}
A. Auerbach, {\em Interacting electrons and quantum magnetism} (Springer, New
  York, 1994).

\bibitem{chen89a}
J.-Q. Chen, {\em Group representation theory for physicists} (World Scientific,
  Singapore, 1989).

\bibitem{lichtenberg70a}
D.~B. Lichtenberg, {\em Unitary Symmetry and Elementary Particles} (Academic
  Press, New York, 1970).

\bibitem{oppermann94a}
O. Veits, R. Oppermann, M. Binderberger, and J. Stein, J. Phys. I France {\bf
  4},  493  (1994).

\bibitem{haldane91a}
F.~D.~M. Haldane, Phys. Rev. Lett. {\bf 67},  937  (1991).

\bibitem{keldysh} L.V.Keldysh.  Sov. Phys. JETP {\bf 20}, 1018 (1965)
 
\bibitem{schwinger} J.Schwinger. J.Math.Phys. {\bf 2}, 407 (1961)

\bibitem{smith} J.Rammer, H.Smith. Rev.Mod.Phys {\bf 58}, 323 (1986)

\bibitem{larkin} M.V.Feigel'man, A.I.Larkin, M.A.Skvortsov. Phys.Rev. 
{\bf B 61}, 12 361 (2000)

\bibitem{kamenev1} A.Kamenev, A.Andreev. Phys.Rev. {\bf B 60}, 2218 (1999)

\bibitem{chamon} C.Chamon, A.Ludwig, C.Nayak. {\it ibid} p.2239 (1999)

\bibitem{kamenev2} A.Altland, A.Kamenev, Phys.Rev.Lett 85, 5615  (2000)

\bibitem{stoof} H.T.C.Stoof, Phys.Rev.Lett {\bf 78}, 768 (1997)

\bibitem{lozano} L.Cugliandolo and G.Lozano. Phys.Rev.{\bf B 59}, 915 (1999)

\bibitem{kree} R.Kree. Z.Phys. B {\bf 65}, 505 (1987)

\bibitem{sompol} H.Sompolinsky. Phys. Rev. Lett {\bf 47}, 935 (1981)
H.Som\-po\-lin\-ski, A.Zip\-pe\-li\-us.  Phys. Rev. Lett {\bf 47}, 359 (1981)

\bibitem{babichenko86a}
V.~S. Babichenko and A.~N. Kozlov. Sol. St. Comm. {\bf 59},  39  (1986).

\bibitem{ioffe} L.B.Ioffe, A.I.Larkin. Phys.Rev. {\bf B 39}, 8988 (1989),

\bibitem{lee} P.A.Lee, N.Nagaosa. {\it ibid}, {\bf B 46}, 5621 (1992)

\bibitem{reizer1} M.Reizer. Phys.Rev. {\bf B 39}, 1602 (1989),
\bibitem{reizer2} M.Reizer. Phys.Rev.  {\bf B 40}, 11571 (1989)

\bibitem{rvb1} G.Baskaran and P.W.Anderson. 
Phys. Rev. {\bf B37}, 580 (1988)

\bibitem{rvb2} G.Baskaran, Z.Zou and P.W.Anderson. 
Sol.St.Commun. {\bf 63}, 973 (1987)

\bibitem{bouis} F.Bouis. {\it These de doctorat de l'ecole polytechnique}. 
Paris (1999)

\bibitem{azakov} S.Azakov, M.Dilaver, A.M.Oztas. Int. Journal of Modern Phys.
{\bf B 14}, 13 (2000), cond-mat/9903379

\bibitem{sachdev1} A.Georges, O.Parcolet and S.Sachdev, 
Phys.Rev.Lett. 85, 840 (2000)

\bibitem{sachdev2} A.Georges, O.Parcolet and S.Sachdev, 
Phys.Rev. {\bf B 63}, 134 406 (2001).

\bibitem{kiselev1} M.Kiselev and R.Oppermann, (unpublished).

\bibitem{kiselev2} M.Kiselev, K.Kikoin and R.Oppermann, cond-mat/0106591.

\bibitem{barz} V.Barzykin and I.Affleck, Phys.Rev. {\bf B 57} ,432 (1997)

\bibitem{kikoin1} M.Pustilnik, Y.Avishai and K.Kikoin, Phys.Rev.Lett. {\bf 84}, 1756 (2000)

\bibitem{glaz1} A.Kaminski, Yu.V.Nazarov and L.I.Glazman, Phys.Rev.Lett {\bf 83}, 384 (1999)

\bibitem{glaz2} L.I.Glazman, F.W.J.Hekking and A.I.Larkin, Phys.Rev.Lett. {\bf 83}, 1830 (1999)

\bibitem{kikoin2} K.Kikoin and Y.Avishai, Phys.Rev.Lett. {\bf 86}, 2090 (2001).


\end{thebibliography}

\end{document}